\documentclass[aps,prb,twocolumn,showpacs]{revtex4}
\usepackage{amsmath}
\usepackage{epsfig}
\usepackage{graphicx}

\def\mb#1{\mbox{\boldmath$#1$}}
\def\fig#1{Fig.~\ref{#1}}

\begin{document}
\title{Generation of dc spin current in a narrow channel
with Rashba and Dresselhaus spin-orbit interaction}
\author{Chi-Shung Tang and Yia-Chung Chang}
\affiliation{Research Center for Applied Sciences, Academia Sinica,
Taipei 11529, Taiwan}
\date{\today }

\begin{abstract}
\ We consider a finite range ac-biased front gate acting upon a
quantum channel with Rashba and Dresselhaus spin-orbit interaction
effects. The ac-biased gate, giving rise to a dynamical Rashba
coupling, causes spin-resolved coherent resonant inelastic
scattering. A pure dc spin current is subsequently generated without
accompanying charge current. In the presence of Dresselhaus effect,
the dc spin current is suppressed in the low kinetic energy regime
but is assisted in the high kinetic energy regime.
\end{abstract}


\maketitle

Manipulation of electron spins is achievable with external active
control, which is a central requirement of spintronic devices.
 Of fundamental interest and practical application is the
spin current in the emerging field of spintronics.\cite{Awschalom02}
One of the key issues is how to generate spin current in spintronic
devices.  The standard way is to inject spin polarized current from
a ferromagnetic electrode.\cite{Wolf01}  However, its efficiency is
usually limited by the poor quality of the interface, and it is
accompanied by the charge current. Narrow gap semiconductor
heterostructures offer an efficient control of spins through
intrinsic spin-orbit interaction (SOI). Several approaches utilizing
spin Hall
effects,\cite{Murakami03,Kato04,Sinova04,Bernevig06,Valenzuela06,Sih06}
magnetic fields,\cite{Muccio02,Brataas02,Zhang03,Wang03}
ferromagnetic materials,\cite{Sun03,Zeng03} or optical
excitations\cite{Bhat00,Hubner03,Stevens03} were proposed.

The success of dc front-gate control for the
measurement\cite{Nitta97} of Rashba coupling
strength,\cite{Rashba60} inspired proposals for spin current
generation by nonmagnetic
means.\cite{Pras03,Governale03,Malsh03,Tang05,Wang06}  These
proposals include adiabatic quantum pumping in a quantum dot with
static SOI\cite{Pras03} interfaced with a time-dependent barrier and
a spatially separated Rashba SOI region,\cite{Governale03} and an
ac-biased Rashba-type two-dimensional (2D) disorder
system\cite{Tang05} or quantum channel.\cite{Wang06} It is known
that the translational invariance is broken in the channel direction
due to a spatially localized time-dependent potential, thus allowing
us to explore coherent resonant inelastic scattering and
time-modulated quasi-bound state features.\cite{Tang96}

In addition to the Rashba SOI,\cite{Rashba60} which is caused by the
structure inversion asymmetry (SIA) of the confining potential of
the 2D trapping well, there is also a Dresselhaus
effect\cite{Dresselhaus55} caused by the bulk inversion asymmetry
(BIA)\cite{Dyakonov86} and the interface inversion asymmetry
(IIA).\cite{Vervoort97}  The contributions associated with BIA and
IIA are phenomenologically inseparable.  The Rashba effect is
usually dominant, but the Dresselhaus effect could be also
observable.\cite{Ganichev04}  In this paper, we consider a narrow
channel formed in a high-mobility electronic quantum well by
applying negative bias on the front split gates.  When a finger gate
is deposited above the split gate separated by an insulating layer,
a local time varying Rashba coupling parameter
$\alpha(\mathbf{r},t)$ can be induced by ac-biasing the finger
gate.\cite{Governale03,Wang06}  We shall explore how the interplay
among the static Rashba, the static Dresselhaus, and the dynamical
Rashba SOI effects influences the efficiency of spin current
generation in the absence of source-drain bias.

The electron transport in a narrow channel in the presence of SOI
can be described by the dimensionless Hamiltonian\cite{Tang96}
\begin{equation}
\mathcal{H}_0 = k^2 + \mathcal{H}_{\rm SO}^0 + V_{\rm c}(y),
\end{equation}
where the first term $k^2=k_x^2 + k_y^2$ denotes the kinetic energy
and the third term $V_{\rm c}(y)=\omega_y^2 y^2$ is a potential that
confines the electron in the $\mathbf{y}$ direction. For a narrow
quantum well along [0,0,1] crystallographic direction, the
unperturbed SOI term $\mathcal{H}_{\rm SO}^0$ involving Rashba and
Dresselhaus interaction effects can be described in terms of
$k$-linear form
\begin{eqnarray}
\mathcal{H}_{\rm SO}^0 &=& \mathcal{H}_{\rm R}^0 + \mathcal{H}_{\rm D}^0 \nonumber \\
&=& \alpha_0 \left( \sigma_x k_y - \sigma_y k_x \right) + \beta_0
\left( \sigma_x k_x - \sigma_y k_y \right),
\end{eqnarray}
where $\sigma_i$ ($i$ = \{$x,y,z$\}) are the Pauli matrices and
$\mathbf{k}=(k_x,k_y)$ is the 2D electron wave vector. The
unperturbed Rashba coupling strength $\alpha_0$ is proportional to
the electric field along $\mathbf{z}$ direction perpendicular to the
2D electron gas.  Moreover, the Dresselhaus coupling strength
$\beta_0$ is determined by the semiconductor material and the
geometry of the sample.

For a narrow wire,\cite{Pereira05} the spin-orbit coupling
contributions can be simplified as $\mathcal{H}_{\rm SO}^0 \approx -
\alpha_0 \sigma_y k_x + \beta_0 \sigma_x k_x$.  The right-going
(left-going) eigenfunctions of the unperturbed Hamiltonian in the
subband $n$ are given by
\begin{equation}
\psi_{n\mathbf{k}\sigma}^{R(L)}(\mathbf{r}) =
\exp\left[ik_{n\sigma}^{R(L)}(\mu) x\right]
\varphi_n(y)\,\chi_{\sigma} ,
\end{equation}
where $\sigma=\pm$ labels the two spin branches, $\chi_{\sigma}$ is
the spinor of branch $\sigma$ with two components given by
$e^{i\theta/2}/\sqrt{2}$ and $\sigma e^{-i\theta/2}/\sqrt{2}$ with
$\theta = \arctan(\alpha_0/\beta_0)$.
  In addition, the
wave vectors are defined by $k_{n\sigma}^R(\mu)= \sqrt{\mu
-\varepsilon_n} - \sigma\gamma_0/2$ and $k_{n\sigma}^L(\mu)=
-\sqrt{\mu - \varepsilon_n} - \sigma\gamma_0/2$, where $\mu$ is the
chemical potential, $\gamma_0 = (\alpha_0^2 +\beta_0^2)^{1/2}$, and
$\varepsilon_n$ is the subband threshold, which is shifted from the
bare subband bottom $\varepsilon_n^0 = (2n+1)\omega_y$ by
$-\gamma_0^2/4$. The total Hamiltonian $\mathcal{H} = \mathcal{H}_0
+ \mathcal{H}_{\rm SO}(t)$ contains a dynamical term, induced by the
ac-biased finger gate, which can be written in the form
\begin{equation}
\mathcal{H}_{\rm SO}(t) = - \frac{\alpha_1}{2}\sigma_y\left\{
k_x,\theta\left(L/2 - \left| x \right|\right) \right\}\cos \omega t,
\end{equation}
where $\theta (x)$ is the step function and \{,\} stands for
anti-commutator.

The scattering wave function of the conduction electron incident
from the left reservoir in the spin state $\sigma$ can be obtained
of the form $\Psi_{\sigma}(\mathbf{r},t) =\sum_n \psi_{n\sigma}(x,t)
\varphi_n(y) \chi_{\sigma}$. In the region $x<-L/2$, the
time-dependent wave function along the channel direction is given by
\begin{equation}
\psi_{n\sigma}(x,t) = e^{ik_{n\sigma}^R(\mu)}e^{-i\mu t} + \sum_m
r_{n\sigma}^m e^{ik_{n\sigma}^L(\mu_m)x} e^{-i\mu_m t},
\end{equation}
where $\mu_m \equiv \mu + m\omega$ and $r_{n\sigma}^m$ denotes the
reflection coefficient of the conduction electron in the subband $n$
and photon sideband $m$. In the region $x>L/2$, the wave function is
simply of the form
\begin{equation}
\psi_{n\sigma}(x,t) = \sum_m t_{n\sigma}^m
e^{ik_{n\sigma}^R(\mu_m)x} e^{-i\mu_m t},
\end{equation}
where $t_{n\sigma}^m$ indicates the corresponding transmission
coefficient.  The longitudinal wave function in the time-modulated
region $|x|< L/2$ is given by\cite{Tang96}
\begin{eqnarray}
\psi_{n\sigma}(x,t) &=& \sum_{m^{\prime}p}\left\{
A_{n\sigma}^{m^{\prime}} e^{ik_{n\sigma}^R(\mu_m)x}
J_p \left( k_{n\sigma}^R(\mu_{m^{\prime}}) \frac{\alpha_1}{\omega} \right) \right.
\nonumber \\
&&+ \left.B_{n\sigma}^{m^{\prime}} e^{ik_{n\sigma}^L(\mu_m)x}
J_p \left( k_{n\sigma}^L(\mu_{m^{\prime}}) \frac{\alpha_1}{\omega} \right) \right\}
\nonumber \\
&& \times \exp\left[-i \mu_{m^{\prime}+p} t \right] \, \sigma_p ,
\end{eqnarray}
where $\sigma_p = {\mb 1}$ if $p$ is even, and $\sigma_p =
-\sigma_y$ if $p$ is odd.  Performing the time-dependent mode
matching at $x=\pm L/2$,\cite{Wang06} one can obtain the reflection
and transmission coefficients, $r_{n\sigma}^m$ and $t_{n\sigma}^m$,
at the edges of the time-modulated region.  Similarly, it is easy to
obtain $\tilde{r}_{n\sigma}^m$ and $\tilde{t}_{n\sigma}^m$ for the
conduction electrons incident from the right reservoir.

Summing over all possible scattered propagating modes from both
reservoirs, the net right-going dc spin current can be expressed as
$I_{\rm S}=I^{\uparrow}-I^{\downarrow}$, where
\begin{equation}
 I^{\sigma} = \frac{e}{h} \int dE\,f(E)\,\left[ T_{RL}^{\sigma}
 + T_{LR}^{\bar{\sigma}}\right]
 \label{I}
\end{equation}
with $f(E)$ being the Fermi function in the reservoirs. In addition,
$T_{RL}^{\sigma}=  \sum_n \sum_m^{\prime} |t_{n\sigma}^{m}|^2
v_n^m/v_n^0$ and $T_{LR}^{\bar{\sigma}} = \sum_n \sum_m^{\prime}
 |\tilde{t}_{n\bar{\sigma}}^{m}|^2
v_n^m/v_n^0$, where $v_n^m \equiv (\mu_m - \varepsilon_n)^{1/2}$.
The spin current conservation is maintained due to the suppression
of spin-flip subband mixing.  Since a symmetric marrow channel
configuration gives $T_{LR}^{\sigma} = T_{RL}^{\bar{\sigma}}$, the
net charge current $I_{\rm Q} = I^{\uparrow} + I^{\downarrow}$ is
identically zero, and a pure nonequilibrium spin current is
generated.

The calculations presented below are carried out under the
assumption that the electron effective mass $m^{\ast}=0.036 m_0$,
which is appropriate to the InGaAs-InAlAs interface. The typical
electron density $n_e \sim 10^{12}$~cm$^{-2}$ and $\hbar\alpha_0
\sim 10^{ - 11}$~eV~m.\cite{Nitta97}  We select $\omega_y =0.035$
such that the subband level spacing, $\Delta\varepsilon =
2\omega_y$, is 4.13 meV.  Accordingly, the length unit $L^{\ast} =
4.0$~nm, the energy unit $E^{\ast} = 59$ meV, and the spin-orbit
coupling parameters are in units of $v_F^{\ast}/2 = 1.8\times 10^5$
m/s.

\begin{figure}[tbhq]
      \includegraphics[width=0.45\textwidth,angle=0]{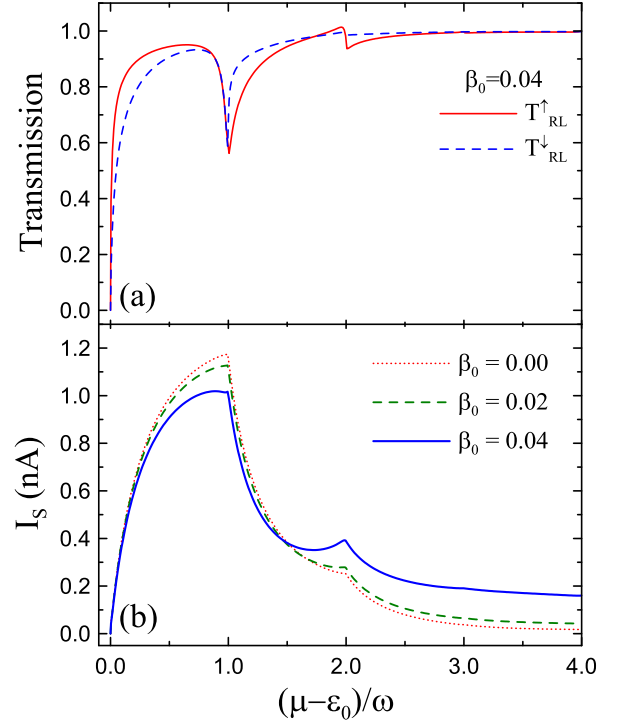}
      \caption{(Color online) (a): Spin-resolved current transmission
      $T_{RL}^{\uparrow}$ (solid red) and $T_{RL}^{\downarrow}$
      (dashed blue) as a function of electron energy in units of driving frequency.
      (b): Generated dc spin current for the cases of
      $\beta_0 = 0.0$ (dotted red), $0.02$ (dashed green), and $0.04$ (solid blue).
      $L/L^{\ast}=30$, $\alpha_0=0.13$, $\alpha_1 =0.05$, and
      $\omega = 0.05 \Delta\varepsilon$.
      }
      \label{fig1}
\end{figure}
In \fig{fig1}, we demonstrate how the Dresselhaus spin-orbit
coupling strength influence the dc spin current generated by the
ac-biased front gate with driving region $L=120$~nm. The other
parameters are static Rashba parameter $\hbar\alpha_0=1.5\times
10^{-11}$~eV~m, $\alpha_1 = 0.38 \alpha_0$, and $\hbar\omega =
0.2$~meV.  In \fig{fig1}(a), we see that since $T_{RL}^{\uparrow} >
T_{RL}^{\downarrow}$ in the low kinetic energy regime
($K\equiv\mu-\varepsilon_0<\omega$), and hence positive spin current
is generated. On the other hand, the sharp dip structure at $K
\approx \omega$ is the one-photon quasibound-state
feature.\cite{Tang96}  For electron energies at $1<K/\omega<2$, we
see clearly $T_{RL}^{\uparrow} < T_{RL}^{\downarrow}$ leading to
positive spin current. The change in sign in the transmission
difference $\Delta T_{RL} = T_{RL}^{\uparrow} - T_{RL}^{\downarrow}$
across the dip structures, that is, $\Delta T_{RL}(K = \omega^{-})>
0$ while $\Delta T_{RL}(K = \omega^{+}) < 0$. Consequently, for
electrons with incident energy $K/\omega \approx 1$, the spin
current peak is generated with the order of 1~nA, as is shown in
\fig{fig1}(b).

For the cases of zero and weak Dresselhaus SOI such as $\beta_0 =
0.0$ and $0.02$,  the electrons with energy $K/\omega \approx 2$
exhibit small dip structures which is associated with two-photon
quasibound-state feature. Since at $K/\omega \approx 2$ the current
transmission $T_{RL}^{\uparrow}$ of spin-$\uparrow$ electron is
still less than $T_{RL}^{\downarrow}$ of the spin-$\downarrow$
electron, there is no significant contribution to the generation of
dc spin current.  In \fig{fig1}(a), we show the current transmission
for the case of strong Dresselhaus coupling $\beta_0 =0.04$.  The
right-going spin-$\uparrow$ electron manifests Fano-type
peak-and-dip line shape in transmission at $K\approx 2\omega$, which
is associated with the two-photon quasibound-state feature. This
Fano-type feature enhances $T_{RL}^{\uparrow}$ to be greater than
$T_{RL}^{\downarrow}$. Consequently, the pumped dc spin current is
thus significantly enhanced.

\begin{figure}[tbhq]
      \includegraphics[width=0.45\textwidth,angle=0]{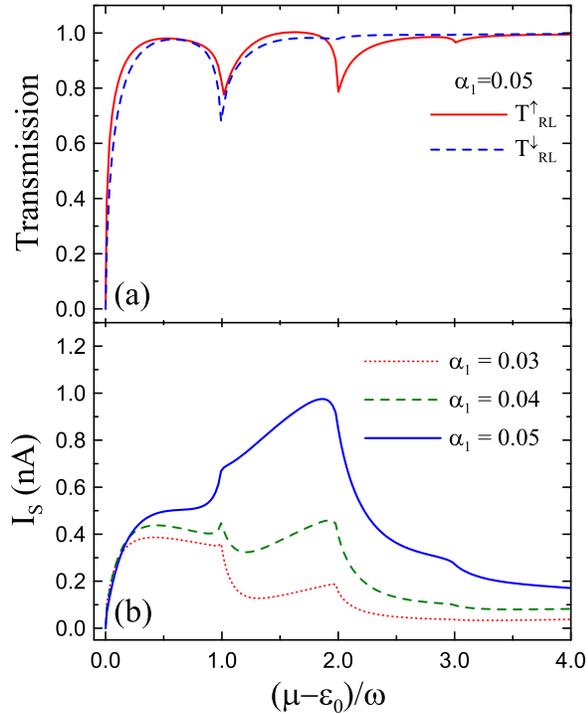}
      \caption{(Color online) (a): Spin-resolved current transmission
      $T_{RL}^{\uparrow}$ (solid red) and $T_{RL}^{\downarrow}$
      (dashed blue) as a function of electron energy in units of driving frequency.
      (b): Generated dc spin current for the cases of
      $\alpha_1 = 0.03$ (dotted red), $0.04$ (dashed green), and $0.05$ (solid blue).
      $L/L^{\ast}=50$, $\alpha_0=0.13$, $\beta_0 =0.03$, and $\omega = 0.05 \Delta\varepsilon$.
      }
      \label{fig2}
\end{figure}
Figure \ref{fig2} shows the spin-resolved transmission with
dynamical Rashba coefficient $\alpha_1 = 0.05$ and the generated dc
spin current for cases of $\alpha_1 = 0.03$, $0.04$, and $0.05$. The
other parameters are: ac-driven region $L=200$~nm, Rashba strength
$\hbar\alpha_0=1.5\times 10^{-11}$~eV~m, Dresselhaus strength
$\beta_0 = 0.23 \alpha_0$, and driving frequency
$f=\omega/2\pi=50$~GHz.   There are common transport features for
the three different dynamical coupling strngth.  The overlap of
$I_{\rm S}$ curves for the incident electron energies $K/\omega \leq
0.2$ implies the transport properties is insensitive to the driving
strength in the very low kinetic energy regime.  The similar plateau
features in $I_{\rm S}$ at $K/\omega<1.0$ are caused by the
competition of the current transmission of spin-$\uparrow$ and
spin-$\downarrow$ electrons.

At incident electron energy $(\mu-\varepsilon_0)/\omega\approx1.0$,
there are kink structure in $I_{\rm S}$ for $\alpha_1 =0.03$, peak
structure for $\alpha_1=0.04$, and shoulder structure for
$\alpha_1=0.05$. For the case of $\alpha_1 =0.03$, the current
transmission of the two spin states are almost the same at $K =
\omega^-$ exhibiting plateau feature while the transmission of the
spin-$\uparrow$ electron is smaller than the spin-$\downarrow$ at $K
= \omega^+$ exhibiting strong drop feature in $I_{\rm S}$.  For the
case of $\alpha_1 =0.04$, the crossover in the transmission for two
spin states leads to the peak structure at $K \approx \omega$.
However, for the case of $\alpha_1 =0.05$, we see that
$T_{RL}^{\uparrow}$ is greater than $T_{RL}^{\downarrow}$ until the
crossover at $K/\omega = 1.87$ leading to the shoulder behavior in
$I_{\rm S}$. This feature in combination with the broad dip
structure, associated with electrons emitting two photons to the
subband threshold forming a quasibound state, results in the broad
spin current peak $I_{\rm S}=0.98$~nA at $K/\omega = 1.87$. Small
hill in $I_{\rm S}$ at $K/\omega \approx 3.0$ is barely recognized.
At $K/\omega = 4.0$, the spin currents are nearly saturated to
$I_{\rm S}= 0.04$, $0.08$, and $0.17$~nA for $\alpha_1 = 0.03$,
$0.04$, and $0.05$, respectively.

In this paper we have investigated non-adiabatically how the dc spin
current is generated under the mechanism of SOI using a dynamical
all electrical control on a split-gate-confined narrow channel. We
have demonstrated nontrivial features concerning the spin current
generation mechanism caused by different strength of Dresselhaus
spin-orbit coupling.  These results provide a robust manner of
generating spin current without accompanying charge current.

The spin current generating features have been demonstrated and
illustrated in detail.  It has been found that the Dresselhaus
spin-orbit coupling intends to suppress the efficiency of spin
current generation in the low kinetic energy regime, while the
Dresselhaus effect can enhance the pumped spin current in the high
kinetic energy regime.  Unlike the parametric quantum pumping, in
which two pumping potentials with a phase difference is
needed.\cite{Muccio02} Our proposed spin current generating device
is achievable using a single ac-biased gate, and should be
achievable within recent fabrication capability.

\ \\
The authors acknowledge the financial support by the National
Science Council and the Academia Sinica in Taiwan. C.S.T. is
grateful to inspiring discussions with L.Y. Wang and C.S. Chu, and
the computational facility supported by the National Center for
High-Performance Computing of Taiwan.

\end{document}